# Low Delay Scheduling of Objects Over Multiple Wireless Paths

Kariem Fahmi [1], Douglas J. Leith[1], Stepan Kucera[2], and Holger Claussen[2]

[1]School of Computer Science and Statistics Trinity College Dublin, Dublin 2, Ireland
{*kfahmi,doug.leith*}@*tcd.ie*
[2]Nokia Bell Labs, Dublin 15, Ireland {*stepan.kucera,holger.claussen*}@*nokia-bell-labs.com*

**Abstract**—In this paper we consider the task of scheduling packet transmissions amongst multiple paths with uncertain, time-varying delay. We make the observation that the requirement is usually to transmit application layer objects (web pages, images, video frames etc) with low latency, and so it is the object delay rather than the per packet delay which is important. This has fundamental implications for multipath scheduler design. We introduce SOS (Stochastic Object-aware Scheduler), the first multipath scheduler that considers application layer object sizes and their relationship to link uncertainty. We demonstrate that SOS reduces the 95% percentile object delivery delay by 50-100% over production WiFi and LTE links compared to state-of-the art schedulers. We extend SOS to utilize FEC and to handle the scheduling multiple objects in parallel. We show that judicious priority scheduling of HTTP objects can lead to a 2-3x improvement in page load times.

## 1 INTRODUCTION

While much attention in 5G has been focused on the physical and link layers, it is increasingly being realized that a wider redesign of network protocols is also needed in order to meet 5G requirements. Transport protocols are of particular relevance, with ETSI recently setting up a working group to study next generation protocols for 5G [1] and initiatives such as Google QUIC [2], Coded TCP [3] and the Open Fast Path Alliance [4]. In this paper we consider next generation edge transport architectures of the type illustrated in Figure 1. Traffic to and from client stations is routed via a proxy located close to the network edge (e.g. within a cloudlet). This creates the freedom to implement new transport layer behavior over the path between proxy and clients, which in particular includes the use of multiple wireless paths at the network edge.

Multipath transport protocols have the potential to improve network performance dramatically by utilizing multiple interfaces simultaneously to improve capacity, latency and reliability. Currently, almost all smart phones are equipped with two radio interfaces (WiFi and Cellular) and the ubiquity of WiFi hot-spots and cellular coverage means that the opportunity for multipath aggregation is almost always there. While multiple commercial entities have begun to seize this opportunity [5], building an efficient, low latency multipath transfer mechanism remains highly challenging [6][7][8]. A primary reason for this is that the transmission delay along each path is typically uncertain and time-varying due to queueing, link layer retransmission, the action of congestion control, etc [9]. Packets sent along different paths therefore frequently arrive out of order

*This publication has emanated from research conducted with the financial support of Science Foundation Ireland (SFI) and is co-funded under the European Regional Development Fund under Grant Number 13/RC/2077.*

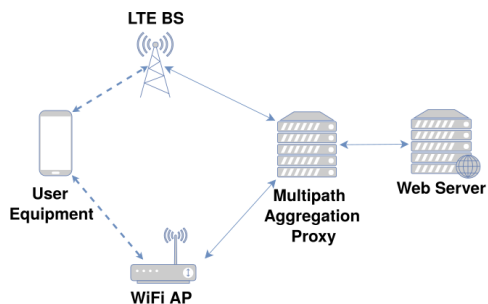

Fig. 1: Schematic of a cloudlet-based edge transport architecture.

and need to be buffered at the receiver to allow in-order delivery, leading to head of line blocking and introducing substantial delay.

In this paper we consider the task of scheduling packet transmissions amongst multiple paths with uncertain, time-varying delay. We make the observation that the requirement is usually to transmit application layer objects (web pages, images, video frames etc) with low latency, and so it is the object delay rather than the per packet delay which is important. This observation fundamentally changes both scheduler design and the scope for making use of links with fluctuating delay. Firstly, in-order delivery of packets is no longer important but rather it is the time when all of the packets forming an object are received which is the key quantity of interest. Secondly, when the requirement is to transmit individual packets subject to a delay deadline then it is difficult to make use of a path with highly fluctuating delay since many packets will miss the deadline. However, when the requirement is to transmit an object consisting of many packets then statistical multiplexing means that it is indeed possible to use a path with fluctuating delay while ensuring low object delivery delay with high probability, as we show in more detail later.

Our main contributions are summarized as follows:

1. We propose SOS (Stochastic Object-aware Scheduler), a multi-path scheduler that takes in consideration the size of the application layer object and the uncertainty of the link delay in order to deliver objects with a reliable bound on delay, making it suitable for latency-sensitive applications. We offer a low complexity implementation that can execute in $O(\log n)$ time for two links, where $n$ is the number of packets in an object. To the best of our knowledge this is the first multipath scheduler that considers application layer object sizes and their relationship to link uncertainty. We demonstrate that SOS reduces the 95% percentile object delivery delay by 50-100% over production WiFi and LTE links compared to state-of-the art schedulers.

2. We extend SOS to utilize FEC (Forward Error Correction) in the form of linear block codes that match the object sizes. By introducing redundancy in this way we show that the scheduler is able to opportunistically exploit periods when the fluctuating path delay is low so as to further reduce object delivery delay, albeit at the cost of slightly reduced throughput capacity.

3. We also extend SOS to handle the scheduling multiple objects in parallel, for example multiple HTTP objects (images, CSS, javascript etc) forming a single web page, while taking into consideration each object's priority. We show that judicious priority scheduling of HTTP objects can lead to a 2-3x improvement in page load times.

## 2 PRELIMINARIES

### 2.1 Application Layer Objects

Applications employ different sizes of objects/frames depending on the type of the application. For example, Fig 2 plots the measured distribution of HTTP object sizes for three popular streaming web sites•. Observe that YouTube video object sizes can be as large as 2500 packets. Similarly, TLS (Transport Layer Security protocol) employs its own framing on top of TCP/UDP which can span up to 13 packets [10] in v1.2, with data unable to be decrypted until a frame is received completely. Compression algorithms such as gzip behave in a similar way.

These observations are pertinent because they mean that the relevant delay for applications is the time it takes to deliver an application object, rather than the time taken to deliver individual packets. Hence, for example, application latency may not be improved by in-order packet delivery (since what matters is that all of the packets forming an object are received, not their order of arrival) and so schedulers that minimize usage of variable-delay links that cause head of line blocking might be inadvertently hurting application latency by reducing aggregate capacity.

---

•Data was collected using HAR files obtained from the Chrome web browser in March, 2018. The object sizes were inferred from the HTTP response content-length field. Youtube video measured is "Ghost Town 8k". NetFlix movie measured is "The Sinner" and a random Twitch stream active at the time was selected.

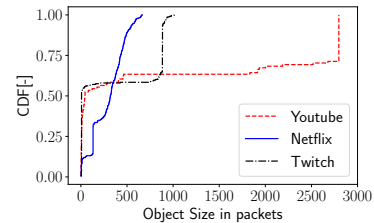

Fig. 2: HTTP object sizes for videos offered by 3 popular streaming websites (Youtube, Netflix and Twitch).

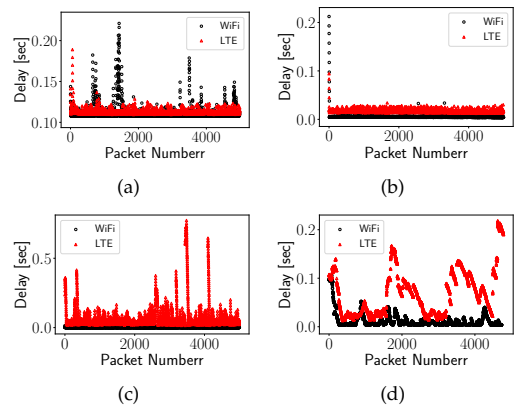

Fig. 3: Packet delay measured while transmitting 5000 UDP packets to a server from 4 different locations in Dublin, Ireland.

### 2.2 Variability of LTE and WiFi Path Delays

Fig. 3 plots example measurements of per packet delay on production WiFi and LTE paths, taken from [9]. These measurements are for UDP packets in order to eliminate the influence of TCP congestion control on the variability of the delay, and are measured between a UE with WiFi and LTE interfaces and a server, both located in the same city. It can be seen that a range of delay behaviours are observed, with the delay behaviour of WiFi and LTE often differing significantly. For example, in Fig. 3(a) it can be seen that the WiFi link is consistently faster than the LTE link (has lower mean delay) but also has higher variability. Conversely, in Figs. 3(c) and 3(d) the LTE path has much higher delay variability than WiFi. The magnitude of the delay fluctuations is also quite variable. For example in Fig. 3(b) the delay fluctuations are around 10-20ms whereas in Fig. 3(c) the LTE delay fluctuations can be as high as 600-800ms.

### 2.3 Multipath Schedulers: State of the Art

Perhaps the most well known multipath scheduler is the MPTCP default scheduler, minRTT [11]. This scheduler attempts to send packets out on the link with the lowest RTT (Round Trip Time) first until its CWND (Congestion Window) is full and then moves to the next lowest RTT link, filling its CWND. Once all CWND are full, it will send packets on a link as soon as there is space in its CWND. Additionally, the scheduler will re-inject packets that are causing HoL blocking on one link into another link and penalize the link that caused the HoL blocking by halving its CWND [11].

In [12][13][14] and [15] various approaches are proposed for reducing out of order delivery in minRTT. In [12], a Blocking Estimation (BLEST) scheduler is introduced. BLEST addresses the issue of HoL blocking when one path is slower than the other, in which it is possible for the faster link to deliver multiple CWNDs worth of packets in the time taken for the slower to link to deliver one CWND worth of packets which, using minRTT, leads to out of order delivery. BLEST estimates the amount of packets that can be delivered on the faster path while the slower path delivers one CWND of packers. It also accounts for growth of the CWND on the faster subflow during the transmission. In [13], a Delay Aware Packet scheduler (DAP) is introduced. DAP creates a schedule based on the Lowest Common Multiple (LCM) of forward delays on all paths. For example, if the forward delay on the slower link is 5 times that of the faster link, then DAP will transmit packets 1 to 5 on the faster link and packet 6 on the slow link. One issue with DAP is that once a schedule is created it will not be modified until completed, which causes it to be less reactive than other approaches. In [14], the Out-of-order Transmission for In-order Arrival Scheduler (OTIAS) is introduced. OTIAS will queue packets to paths regardless of whether they have free space in the CWND. It schedules on a per-packet basis as soon as the packet arrives and creates a queue on each path. In [16], an Earliest Completion First (ECF) scheduler is introduced, which takes into consideration the RTT of each path and decides whether it is faster for a packet to wait for the faster path to have space open in its CWND or to send it on the slower path instead. It also attempts to address path delay variance in a simple manner by calculating the standard deviation of the RTT on each path and adding it to the value of the RTT on each path. One issue with this approach however is that it will not schedule packets out of order and as such if it is better for the current packet to wait for the faster path, other packets waiting in the send buffer will also have to wait.

In [17] and [18] the impact of packet losses is considered, in addition to delay. In [17], a scheme is introduced to avoid scheduling packets on congested paths, aimed at reducing losses. The scheme calculates an estimate for the path capacity based on the value of the CWND half way between the last loss event and the current SSThreshold. Once the number of in-flight packets exceeds this estimate, the estimate is updated to the value of in-flight packets. Furthermore, a threshold is established to identify when the path is congested based on the ratio between the number of in-flight packets and the capacity estimate. A packet is then scheduled on a path if it will not cause it to exceed the congestion threshold. If both flows are below the threshold, the path with the lowest RTT is selected. Essentially, this scheme trades packet delay for reducing the probability of packet loss. In [18], a Fine-grained Forward Prediction based Dynamic Packet Scheduler ($F^2P - DPS$) is introducecd, which attempts to calculate how many CWNDs worth of packets can be transmitted on the faster path while the slower path delivers its CWND, taking into consideration losses. It calculates the expected amount of losses on each link and whether recovery will be done via RTO or Fast Retransmission.

Perhaps the closest work to the present paper in the sense that it takes explicit account of delay variability when making scheduling decisions is [19] , which introduces Stochastic Earliest Deadline Path First (SEDPF). SEDPF models inter-packet delay as i.i.d. Gaussian random variables. Using the Gaussian assumption, it calculates the distribution of the total inter-packet delay of in-flight packets on each link with the new packet added. It then evaluates the maximum of multiple sets of Gaussian random variables, using an approximate closed form, each representing the delay of a packet being sent on a particular path. It then selects the set that had the lowest maximum, as that implies the lowest delay.

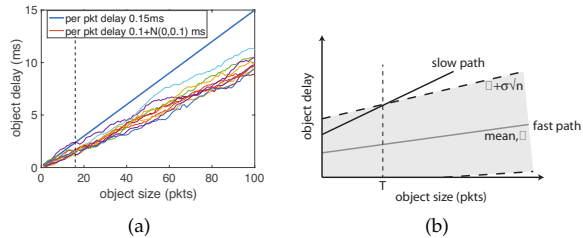

Fig. 4: Illustrating time taken to transmit an object over a slower path with fixed delay and over a faster path with fluctuating delay. Plot (a) shows some example sample paths when the per packet delay on the faster path is normally distributed with standard deviation $\sigma = 0.1$ms. The shaded region in plot (b) indicates schematically the mean plus one standard deviation in transmission time for the faster path. The vertical dashed line on both plots indicates the object size above which the delay on the faster path is, with high probability, lower than that of the slower path.

None of the above schedulers exploit knowledge of application layer objects and their impact on scheduling decisions, which is our focus in the present paper. With the exception of [19], existing schedulers also deal only indirectly with the time-varying and uncertain nature of path delay, namely by utilizing a smoothened average RTT and applying this to all packets sharing the same CWND[§] when making scheduling decisions.

## 3 SCHEDULING OBJECTS

### 3.1 QoS Over Paths With Time-Varying Delay

Consider two paths, one of which is slower but has consistent delay and the other which is faster but has highly fluctuating delay. When the requirement is to transmit individual packets subject to a delay deadline then it is difficult to make efficient use of the link with fluctuating delay since many packets will miss the deadline. However, when the requirement is to transmit an object consisting of multiple packets subject to an overall deadline then the situation changes fundamentally.

The reason why is illustrated in Figure 4. Figure 4(a) plots sample paths of the object transmission time on a fluctuating faster path vs the object size. It can be seen that the variance of the transmission time increases with the object size, as expected since the object transmission time is the sum $\sum_{i=1}^{n} T_i$ of the individual packet transmission times $T_i$ and when these are i.i.d this has standard deviation $\sqrt{n}\sigma$ where $\sigma$ is the standard deviation of the per packet times $T_i$.

---

[§]The difference in delay between the head of the CWND and the tail can be large when network buffers start to fill.

However, the transmission time on the slower path scales with $n$, see Figure 4(b). Hence, for a sufficiently large object size the transmission time will, with high probability, be lower on the faster path despite its highly fluctuating delay.

This basic observation has significant implications. In particular, it means that we can still make efficient use of wireless links with highly fluctuating per packet delay (e.g. mmWave, LiFi) while providing controlled quality of service, provided that when scheduling packets across paths we move from consideration of packets individually to consideration of objects (collections of packets subject to an overall delivery deadline).

## 3.2 Object Scheduler Design

In light of the above observation we would like to design a multipath packet scheduler that is cognisant of (i) objects within the packet stream and (ii) the impact of fluctuations in path delay on object transmission time.

Index the available paths by $j = 1, 2, \ldots, m$ and let $T_k^{(j)}$ and $P^{(j)}$ denote the inter-packet delay and propagation delay, respectively, experienced by packet $k = 1, 2, \ldots$ transmitted on path $j$. Given an object consisting of $n$ packets to be transmitted across the paths let $n^{(j)}$ denote the number of packets sent over path $j$, with $\sum_{j=1}^{m} n^{(j)} = n$, and $\vec{n} = [n^{(1)}, \ldots, n^{(m)}]^T$. The transmission time of the object is then given by

$$D(\vec{n}) := \max_{j \in \{1, \ldots, m\}} T^{(j)}(n^{(j)}) + P^{(j)} \quad (1)$$

where $T^{(j)}(n^{(j)}) = \sum_{k=1}^{n^{(j)}} T_k^{(j)}$. The difficulty in designing a scheduler is, of course, that the per packet delays $T_k^{(j)}$ are variable and unknown, and so also the aggregate delays $T^{(j)}(n^{(j)})$. While we might attempt to calculate the probability distribution of transmission time $D(\vec{n})$ for every partition $\vec{n}$ this quickly becomes computationally expensive for larger object sizes (so infeasible for real-time packet scheduling) plus in any case we usually lack full details of the distribution of the per packet delays on each path. We therefore adopt the following approximate approach and select parameters $w^{(j)} \geq 0$ such that

$$P(T^{(j)}(n^{(j)}) \geq T_U^{(j)}(n^{(j)})) \leq \epsilon^{(j)} := \epsilon/m \quad (2)$$

where $\mu^{(j)} = E[T_k^{(j)}] \geq 0$, $T_U^{(j)}(n^{(j)}) = n^{(j)}\mu^{(j)} + \sqrt{n^{(j)}}w^{(j)}$ and $\epsilon$ is a design parameter which, unless otherwise stated, in the rest of the paper we select to be 0.05. In particular, modelling the inter-packet delays $T_k^{(j)}$ as being i.i.d then using the Chernoff-Hoeffding bound we can solve for $w^{(j)}$ using

$$w^{(j)} = \sqrt{\frac{-\ln(\epsilon^{(j)})(a^{(j)} - b^{(j)})^2}{2}} \quad (3)$$

where $a^{(j)}$ and $b^{(j)}$ are upper and lower bounds on the inter-packet delay, $a^{(j)} \leq T_k^{(j)} \leq b^{(j)}$. It follows from (2) that

$$P(D(\vec{n}) \geq D_U(\vec{n})) \leq 1 - (1 - \epsilon/m)^m \leq \epsilon \quad (4)$$

where $D_U(\vec{n}) = \max_{j \in \{1, \ldots, m\}} T_U^{(j)}(n^{(j)}) + P^{(j)})$. We now select a partition $\vec{n}$ that solves the following optimization problem,

$$\min_{\vec{n} \in \mathbb{N}^m} \quad D_U(\vec{n}) \quad s.t. \quad \sum_{j=1}^{m} n^{(j)} = n \quad (5)$$

### 3.2.1 Solving (5)

When we relax the optimisation (5) so that $\vec{n} \in \mathbb{R}_+{}^m$ i.e. a fractional number of packets can be sent on each path, then for sufficiently large objects the solution to the optimisation ensures equal 95th percentile delay on all paths and so is of the Wardrop form. That is, we have:

**Lemma 1** (Wardrop Optimum). *Suppose that for each path $j = 1, \ldots, m$ the mean path delay $\mu^{(j)}$ and variability parameter $w^{(j)}$ are not both zero i.e. the inter-packet delay is not identically zero on any path. Then the solution $\vec{n}^*$ to relaxed optimisation $\min_{\vec{n} \in \mathbb{R}_+{}^m} \quad D_U(\vec{n}) \quad s.t. \quad \sum_{j=1}^{m} n^{(j)} = n$, satisfies*

$$|T_U^{(j)}((n^*)^{(j)}) - T_U^{(i)}((n^*)^{(i)}) + P^{(j)} - P^{(i)}| = 0 \quad (6)$$

*for all $i, j \in \{1, \ldots, m\}$ provided $n$ sufficiently large that all elements of $\vec{n}^*$ are non-zero i.e. all paths are used to transmit packets.*

*Proof.* We proceed by contradiction. Suppose $\vec{n}^*$ is optimal and $D_U(\vec{n}^*) = T_U^{(i)}((n^*)^{(i)}) + P^{(i)} + \delta$, $\delta > 0$, for some $i$. Let $L = \{j \in \{1, \ldots, m\} : D_U(\vec{n}^*) = T_U^{(j)}((n^*)^{(l)}) + P^{(l)}\}$ i.e. the set of paths with maximal delay. We can always decrease $(\vec{n}^*)^{(j)}$ by $\epsilon > 0$ for $l \in L$ and increase $(n^*)^{(i)}$ by $\epsilon|L|$ so that the constraint $\sum_{j=1}^{m}(n^*)^{(j)} = n$ remains satisfied. Select $\epsilon$ such that $((n^*)^{(i)} + \epsilon|L|)\mu^{(i)} + w^{(i)}\sqrt{(n^*)^{(i)} + \epsilon|L|} + P^{(i)} \leq \max_{l \in L}((n^*)^{(l)} - \epsilon)\mu^{(l)} + w^{(l)}\sqrt{(n^*)^{(l)} - \epsilon} + P^{(l)}$. This is always possible since $(n^*)^{(i)}\mu^{(i)} + w^{(i)}\sqrt{(n^*)^{(i)}} + P^{(i)} + \delta = (n^*)^{(l)}\mu^{(l)} + w^{(l)}\sqrt{(n^*)^{(l)}} + P^{(l)}$, by assumption, and $(n^*)^{(i)} > 0$. Since one of $\mu^{(j)}$ and/or $w^{(j)}$ is greater than zero this change decrease the maximum path delay, yielding the desired contradiction. □

When we now restrict ourselves to sending an integer number of packets on each path essentially the same reasoning tells us that the solution to optimisation (5) ensures approximately equal 95th percentile delay on all paths, with inequalities in delay arising due to quantisation. That is, we can solve optimisation (5) by first solving the convex optimisation in Lemma 1 and then searching over the integer vectors obtained by taking the ceil or floor of each element of the convex solution to find the integer vector minimising $D_U$. This search is over $2^m$ combinations and so is fast for realistic values of $m$ e.g. $m \leq 4$. Alternatively, when $m = 2$ then the solution to (5) can also be found by bisection search with time complexity of $O(\log n)$.

### 3.2.2 Estimating $E[T_k^{(j)}]$, $a^{(j)}$ and $b^{(j)}$

The mean $E[T_k^{(j)}]$, lower bound $a^{(j)}$, and upper bound $b^{(j)}$ of the inter-packet delay on path $j$ are estimated as, respectively, the average, minimum and 95% percentile of the last 5000 packets, or all previous packets if fewer than 5000 were sent (5000 is used as it is a reasonable upper bound on object size). The inter-packet delay is calculated by observing the time between arrival of acknowledgements at the sender. This way we allow the scheduler to be implemented without any modifications to the receiver. Note also that with this approach the link parameters are updated on every ACK, allowing the scheduler to be react quickly to changes.

### 3.2.3 Receive Buffer Management

Packets are buffered at the receiver until they can be delivered in-order to the application so as to preserve TCP byte-stream semantics. This means that the receiver can become blocked when the receive buffer capacity is reached due to out-of-order delivery or outage on one or more paths. To reduce the occurrence of such blocking we size the receive window as $\sum_{j=0}^{m} \frac{D_U(\vec{n})}{\mu^{(j)}}$. In addition, in the rare cases where blocking does occur we re-transmit the packets which fail to be delivered.

### 3.2.4 Multiple Objects

When multiple objects need to be scheduled, or one object arrives before the previous object is fully transmitted, then the expression for $T_U^{(j)}$ is modified in the obvious way to:

$$T_U^{(j)}(n^{(j)}) = (n^{(j)} + u^{(j)})\mu^{(j)} + (\sqrt{n^{(j)} + u^{(j)}})w^{(j)} \quad (7)$$

where $u^{(j)}$ is the number of packets in-flight on link $j$. Additionally, when scheduling multiple objects belonging to different connections, SOS can prioritize more important objects through changing the order of sending or preemption.

## 3.3 Performance Evaluation

In this section we evaluate the performance of the proposed object scheduler, which we refer to as SOS (Stochastic Object-aware Scheduler) across a range of link specs and object sizes. To provide a baseline we compare this with the performance of the Earliest Deadline First (EDF) and Stochastic Earliest Deadline Path First (SEDPF) packet schedulers, EDF being optimal when path delays are fixed and known and SEDPF being a state of the art multipath packet scheduler introduced in [19] that takes explicit account of path delay fluctuations.

### 3.3.1 Synthetic Delay Data

We begin by evaluating performance across two paths where the inter-packet delays are synthetically generated and follow a Gamma distribution (we note that [20], [21] observe that network delay tends to follow a Gamma distribution with a heavy tail). In the plots L1(1,$\sigma^2$) indicates that path one has mean inter-packet delay 1ms and that the standard deviation $\sigma$ is marked on the x-axis of the plot. Similarly, L2 indicates the properties of path two. The paths specs are selected to illustrate the performance both when having very stable links and highly variable links. Additionally, we alternate between the faster link being more variable and the slower link more variable. For each scheduler the mean delay and the standard deviation are provided to each scheduler so as to eliminate the estimation of the channel as a variable in the results.

Figure 5 compares the performance of the SOS and SEDF schedulers. In left-hand part of Figures 5(a)-(d) path one has mean delay 10ms and standard deviation 1ms while path two has mean delay 12ms and standard deviation is varied as indicated on the x-axis. That is, path two is slightly slower on average than path one. The right-hand side of the plot shows the performance when path two has its standard deviation held fixed and 1ms and the standard deviation of path one is varied as indicated on the x-axis. Figure 5(a) shows results for objects consisting of a single packet, and Figure 5(b)-(d) show results as the object size is increased to 10, 100 and 1000 packets respectively.

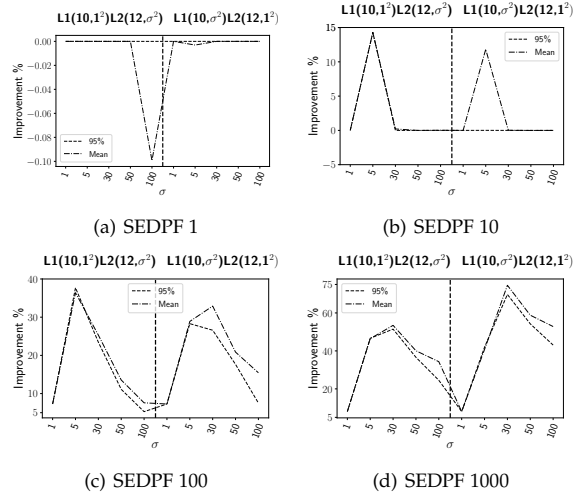

(a) SEDPF 1      (b) SEDPF 10

(c) SEDPF 100      (d) SEDPF 1000

Fig. 5: Comparing SOS against SEDPF. Plot shows improvement by SOS at both 95th percentile and mean over synthetic links generated using Gamma distributed inter-packet delays.

For objects consisting of a single packet it can be seen from Figure 5(a) that SOS and SEDPF have essentially identical performance. In this case both schedulers opt not to use the more variable link. However, it can be seen from Figures 5(b)-(d) that as the object size increases, SOS begins to achieve increasingly large improvements in both the mean and 95th percentile delay over SEDPF. What is happening is that SOS utilizes the variable link faster than SEDPF as it is aware that the larger object size will ensure less variability.

Figure 6 shows the corresponding results comparing SOS and EDF. On the RHS of Figure 6(a) it can be seen that even for objects consisting of a single packet SOS achieves substantial improvements (of around ×3.5) over EDF in the 95th percentile delay. On the LHS both schedulers perform similarly since when they both opt to use the faster path when the slower path is more variable. Note that the 95th percentile is roughly an upper bound on the delay and so is often the quantity of most interest for latency sensitive applications (VOIP, video chat, gaming) where objects are required to be delivered within a specified deadline. Figure 6(b)-(d) also show large improvements in the 95th percentile delay, although it can be seen that these come at the cost of reduced mean delay performance. That is, SOS trades off mean delay performance for improved worst-case performance, as expected.

In summary, by being cognizant both of path delay fluctuations and that it is the time to deliver objects rather than packets which is the quantity of interest, SOS offers significant gains in delay performance over state of the art packet schedulers.

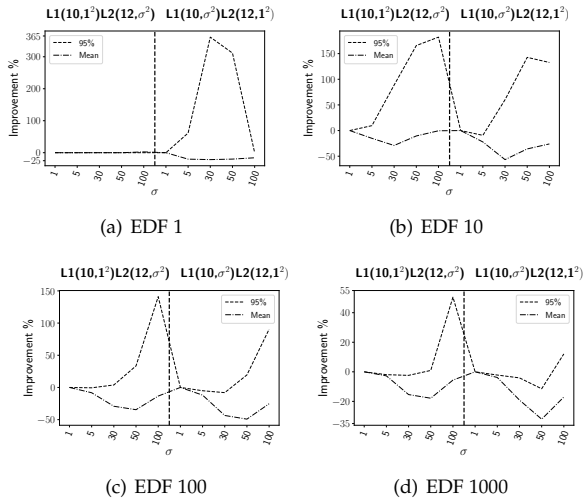

Fig. 6: Comparing SOS against EDF. Plot shows improvement by SOS at 95th percentile and mean delay over synthetic links generated using Gamma distributed delays.

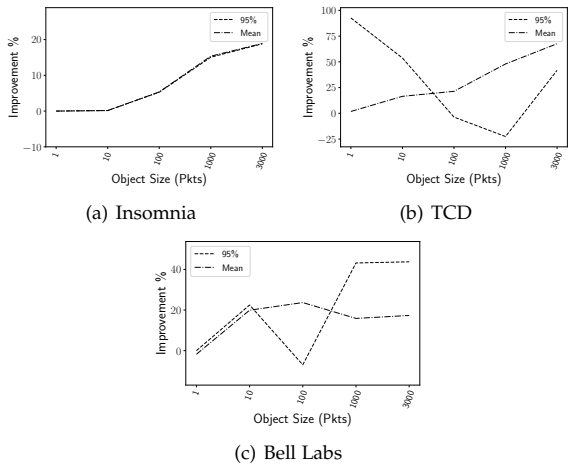

Fig. 7: Comparing SOS against SEDPF. Plot shows improvement by SOS at 95th percentile and mean delay over experimentally measured packet delay data collected at 3 locations.

### 3.3.2 Experimental Measurements

In addition to the above evaluation using synthetic delays we also collected packet delay data in three different locations in Dublin, Ireland, namely a Coffeeshop (Insomnia), the Nokia Bell Labs site and the Trinity College Dublin campus. The trace contains inter-packet delay measured simultaneously over two parallel TCP connections (one for WiFi and one for LTE) using the default CUBIC congestion control between a laptop and an AWS Linux box. The LTE connection was created through USB tethering to a smart-phone. The delay was measured using Wireshark timestamps and 20000 packets were sent over each link at each location. Using this trace data we then measured the performance of the SOS, EDF and SEDPF schedulers.

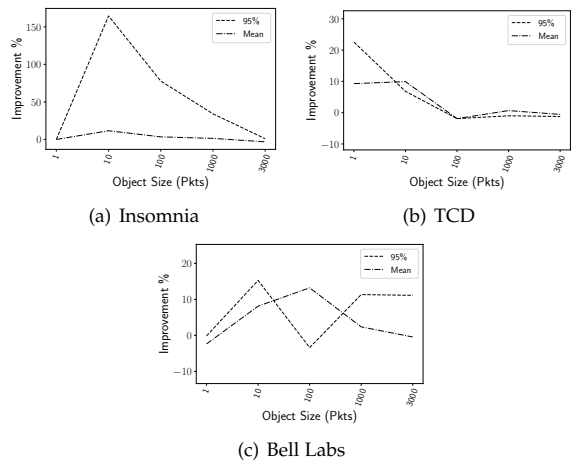

Fig. 8: Comparing SOS against EDF. Plot shows improvement by SOS at 95th percentile and mean delay over experimentally measured packet delay data collected at 3 locations.

Figure 7 compares the measured 95th percentile and mean delay performance of the SOS and SEDPF schedulers at the three measurement locations as the object size is varied. Figure 7(a) shows the same general trend as Figure 5, with SOS offering improvements in both the 95th percentile and mean delays over SEDPF. Figure 7(c) shows similar behavior, apart from a small dip in the 95th percentile improvement for object size 100 packets. In Figure 7(b) the behaviour is somewhat different, with the improvement in the 95th percentile delay greatest for smaller object sizes while the improvement in the mean delay increases with object size. This is due to a few extreme outliers in the packet delays at this location that heavily influenced the standard deviation, and resulted in SEDPF staying clear of the WiFi link, despite the fact that it had lower 95th percentile delay. In SOS, we don't consider the standard deviation, but rather the 95th percentile delay and the minimum, and as such it is more resilient to outliers.

Figure 8 shows the corresponding results comparing SOS and EDF. It can be seen that the 95th percentile performance behaves similar to that in Figure 6, but unlike in Figure 6 this does not come at the cost of reduced mean delay performance.

## 4 OPPORTUNISTICALLY LOWERING LATENCY

### 4.1 Motivating Example

The SOS scheduler introduced in Section 3.2 allocates packet transmissions for an object amongst the available paths so as to minimise the 95th percentile delay, thereby ensuring that an object is with high probability delivered with low delay. However, we can also sometimes be "lucky" in the sense that the packet delays actually realised on one or more paths happen to be lower than usual.

For example, suppose we have two paths with 95th percentile delays $T_U^{(1)}(n)$ and $T_U^{(2)}(n)$ as indicated in Figure 9 and the specific random delay realisation $T^{(2)}(n)$ for path 2 corresponding to the lower line in the figure. Suppose for simplicity also that the path 1 delay is deterministic and so the delay realisation $T^{(1)}(n) = T_U^{(1)}(n)$ i.e. lies on the dashed line in Figure 9. For an object consisting of $n < n_0$ packets it can be seen that to minimise the 95th percentile delay we should send all $n$ packets on path 1. However, in this example we are lucky and happen to see a sequence of low delays on path 1 so that $T^{(2)}(n) < T^{(1)}(n)$ for all $n < n_0$. That is, if we had instead sent all $n$ packets on path 2 rather than path 1 then we would have achieved a lower delay. Note that a different realisation of the random delays, such as that marked $T^{(2)}(n)'$ on the figure, might lead to scheduling on path 1 yielding the lower delay.

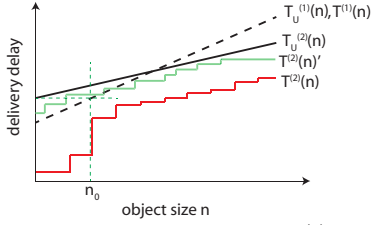

Fig. 9: Illustrating how a "lucky" realisation $T^{(2)}(n)$ of the random delay on a path (the lower line) can potentially allow scheduling delay to be reduced compared to a more "typical" realisation $T^{(2)}(n)'$.

Of course we only discover the delay realisation on a path *after* transmitting our packets. Nevertheless, by transmitting additional redundant (coded) packets, over and above the minimum $n$ required to communicate an object, we can still opportunistically reduce the object delivery delay by exploiting path low delay realisations when they occur. For example, suppose we construct as many redundant packets as required by using a random linear code[†] such that, roughly speaking, on receipt of any $n$ coded and uncoded packets then the object can be reconstructed at the receiver. Considering again the example in Figure 9, we now schedule $n$ packets on path 1 and $n$ on path 2 i.e. $2n$ packets in total. Then we still minimise the 95th percentile delay, as before, since we send $n$ packets down path 1. However, by sending additional packets down path 2 then if we are lucky and experience a delay realisation as in the figure then we are able to take advantage of this to achieve a lower object delivery time. Observe that in this way we are trading off network capacity (since we send more packets) for the chance of lower delay.

### 4.2 SOS-FEC Opportunistic Scheduler

To take advantage of path realisations with low delay we modify the SOS scheduler from Section 3.2 to send $n + \delta$ packets, $\delta \geq 0$, where receipt of any $n$ of the $n + \delta$ packets sent is sufficient to allow the object to be reconstructed at the receiver. On each path $j = 1, \ldots, m$ we send $n^{(j)} + \delta^{(j)}$ where $n^{(j)}$ is selected as in Section 3.2 and $\sum_{j=1}^{m} \delta^{(j)} = \delta$. It remains to decide the value $\delta$ for the number of additional packets sent and how these additional $\delta$ packets are to be distributed across the available paths i.e. how to select the $\delta^{(j)}$, $j = 1, \ldots, m$.

---

[†] In more detail, we treat the packets forming an object as symbols in an appropriate finite field and construct coded packets as weighted linear combinations of these symbols where the weights are selected uniformly at random from the finite field.

For each path $i = 1, \ldots, m$ we solve the following optimization

$$\vec{\eta}^* \in \arg \min_{\vec{\eta} \in \mathbb{N}^m} \quad D_U^{(i)}(\vec{\eta}) \quad s.t. \quad \sum_{j=1}^{m} \eta^{(j)} = n \qquad (8)$$

where $D_U^{(i)}(\vec{\eta}) = \max_{j \in \{1, \ldots, m\}} T_U^{(i,j)}(\eta^{(j)}) + P^{(j)}$ and

$$T_U^{(i,j)}(\eta^{(j)}) = \begin{cases} \eta^{(j)} \mu^{(j)} + \sqrt{\eta^{(j)}} w^{(j)} & j \neq i \\ \eta^{(i)} \mu^{(i)} + \gamma \sqrt{\eta^{(j)}} w^{(j)} & otherwise \end{cases} \qquad (9)$$

with $0 \leq \gamma < 1$ a design parameter. We then select $n^{(i)} + \delta^{(i)} = (\eta^*)^{(i)}$. That is, we select the number $n^{(i)} + \delta^{(i)}$ of packets to send on path $i$ to be equal to the $i$'th element of the optimal split vector $\vec{\eta}^*$ when the delay fluctuations on path $i$ are a factor $\gamma$ better than the 95th percentile. We refer to this scheduler as SOS-FEC (since it uses FEC to generate the $\delta$ redundant packets).

This approach has a number of appealing properties. Firstly, we note that it always selects $\delta^{(i)} \geq 0$ since $\gamma < 1$ reduces the delay $T_U^{(i,i)}$ in (9) compared to $T_U^{(i)}$ and this will increase the number of packets assigned to path $i$. Secondly, while we might generalise (9) to

$$T_U^{(i,j)}(\eta^{(j)}) = \eta^{(j)} \mu^{(j)} + \gamma^{(i,j)} \sqrt{\eta^{(j)}} w^{(j)} \qquad (10)$$

with $0 \leq \gamma^{(i,j)} \leq 1$ with a view to exploiting combinations of low delay realisations in multiple paths, in fact the solution to (9) will always dominate the solution to (10) when $\gamma^{(i,i)} = \gamma$ (for consistency with (9)). To see this observe that selecting $\gamma^{(i,j)} < 1$ for paths $j \neq i$ will tend to increase the number of packets assigned to path $j$. Since the total number of packets $\sum_{j=1}^{m} \eta^{(j)} = n$ in (8), this means that the number of packets assigned to path $i$ will tend to decrease. Hence, (9) yields packet allocation to each path which is at least as large as would be assigned by (10).

### 4.3 Performance Evaluation

#### 4.3.1 Varying the Gamma parameter

In Fig. 10, we measure the performance of the SOS-FEC scheduler using a range of values for the $\gamma$ parameter. The inter-packet delay distribution for the two links and the object size is kept static across the tests. As expected the amount of redundancy is reduced as the value of $\gamma$ goes to 1, where the amount of redundancy becomes 0 at $\gamma = 1$. There's more significant improvement in the mean compared to the 95th% percentile, as the mean improvement increases by 75% vs SEDPF and 50% vs EDF across the range of $\gamma$ values used, while the 95% percentile improvement is approximately 25% and 35% vs EDF and SEDPF respectively. This is because the scheduling decisions of SOS are already optimized for the 95% percentile and as such the improvement due to FEC would mostly arise if the instances of "unlucky" realizations of delay on one paths occur simultaneously with the "lucky" realizations on the other paths. The value of the redundancy $\delta$ increased by roughly 35% as the value of the $\gamma$ parameter varied from 1 to 0.

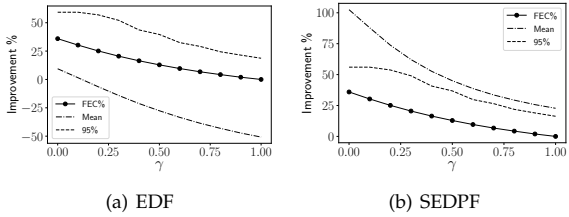

Fig. 10: Comparing SOS-FEC against EDF and SEDPF: plot shows improvement at 95% percentile, mean delay and amount of redundancy $\delta$ across a range values of the $\gamma$ parameter. The inter-packet delay of the two paths is Gamma distributed with means $\mu_1 = 10ms, \mu_2 = 12ms$ and standard deviations $\sigma_1 = 50ms, \sigma_2 = 1ms$ The object size is 100 packets.

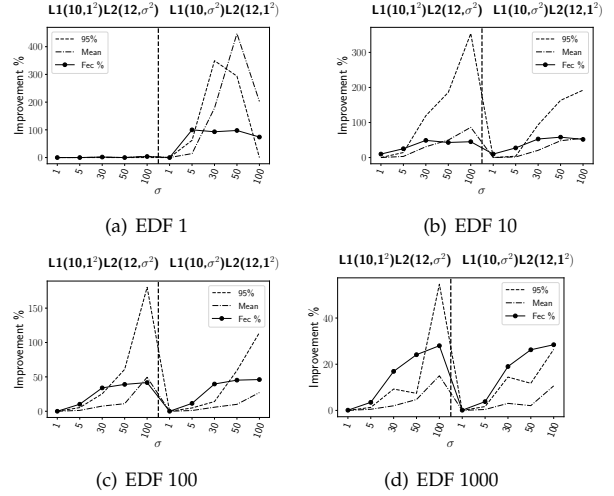

Fig. 12: Comparing SOS-FEC against EDF. Plot shows improvement by SOS at mean and 95th percentile delay, and also the level of redundancy $\delta$.

### 4.3.3 Experimental Measurements

Figure 13 compares the delay performance of SOS-FEC and SEDPF for packet delay traces measured at three locations in Dublin, Ireland i.e. the same delay measurements as used in Section 3.3.2. It can be seen that, compared to SOS, SOS-FEC achieves a larger improvement in both mean and 95th percentile delays for smaller object sizes. This is because for smaller object sizes tends to SOS stay clear of variable paths even when they that have significantly lower mean delay whereas SOS-FEC is able to utilize those paths without compromising the 95th percentile delay. Similarly to the synthetic data, the redundancy decreases as the object size increases, until it levels off at 3000 packet objects where there's almost no redundancy.

Figure 14 shows the corresponding delay performance of SOS-FEC and EDF. The general improvement pattern as object size increases is similar to SOS vs EDF i.e. it levels off as the object size increases. However, unlike SOS, we can observe significant improvement in the mean delay. This is because EDF makes poor decisions due to outliers influencing the mean delay estimated using a rolling window. However, as the object size increased it can be seen that the performance of EDF and SOS-FEC becomes much the same as they both converge to similar scheduling decisions.

## 5 OBJECT PRIORITIZATION

To take advantage of another aspect of object-aware scheduling, we evaluate the potential performance gains when using SOS, with knowledge of object priorities, to download the objects forming web pages we collected the HTTP object sizes for the pages from the home pages of the 13 most visited websites in Ireland according to the Alexa.com ranking, using a custom addon for the chrome browser. The data was collected on March,2018. Using this addon we also collected the sequence of object requests, the initiator tree, the connection id of the object, and the subset of objects required for the DOM rendering to complete i.e. the event after which the website is visible to the user. It is

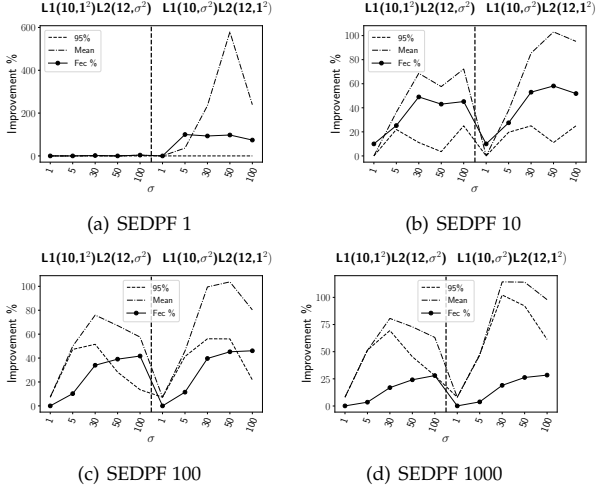

Fig. 11: Comparing SOS-FEC against SEDF. Plot shows improvement by SOS at mean and 95th percentile delay, and also the level of redundancy $\delta$.

### 4.3.2 Synthetic Delay Data

As before, we begin by evaluating performance using synthetic Gamma distributed delay data.

Fig. 11 compares the delay performance of SOS-FEC and SEDPF for a range of path conditions and object sizes. For objects consisting of a single packet it can be seen from Fig. 11(a) that when the faster link has more variable delay (the RHS of the plot) both the mean and 95th percentile delays are improved by a factor of 300-400% when using SOS-FEC. Of course, this comes at the cost of added redunancy, in this case sendign roughly twio packets instead of a single packet for each object. For larger objects it can be seen from Fig. 11(b)-(d) offers substantial gains in 95th percentile delay and also smaller, but still significant, gains in mean delay across a wide range of path conditions. It can also be seen that the amount of redundancy used by SOS-FEC scales with the path delay variability, as might be expected, and also tends to decrease with increasing object size.

Fig. 12 shows the corresponding result for EDF. It can be seen that once again SOS-FEC achieves substantial improvements in the 95th percentile delay across a wide range of path conditions and object sizes.

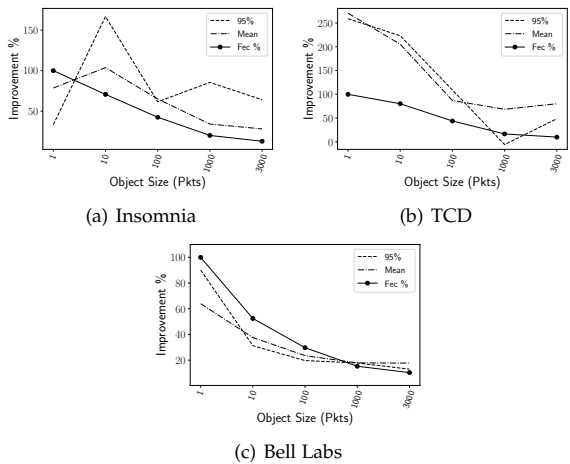

(a) Insomnia     (b) TCD

(c) Bell Labs

Fig. 13: Comparing SOS-FEC against SEDPF. Plot shows improvement by SOS at 95th percentile and mean delay, and the level of redundancy, over experimentally measured packet delay data collected at 3 locations.

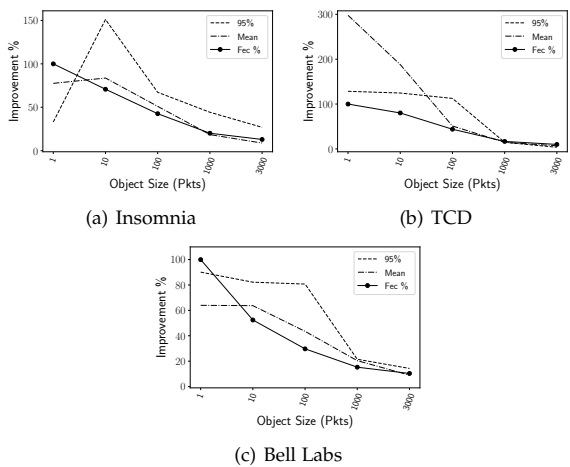

(a) Insomnia     (b) TCD

(c) Bell Labs

Fig. 14: Comparing SOS-FEC against EDF. Plot shows improvement by SOS at 95th percentile and mean delay, and the level of redundancy, over experimentally measured packet delay data collected at 3 locations.

the delay to the rendering complete event that is typically of most impact on user quality of experience.

Figure 15(a) shows an example web page transmission where the page had 3 objects, 2 of which belong to the DOM which means they are needed before rendering of the page can begin. The first packet of the HTML object generated a request for a non-DOM object and the second packet generated a request for a DOM object. The transmission of the objects follows the order of the requests, which resulted in a slower rendering time of the page as the non-non-DOM object was queued for transmission ahead of the DOM object. Figure 15(b) shows the division of DOM and non-DOM objects in a website. It can be observed that on average, DOM objects are a small fraction of the total objects in a website.

To address the problem of in-optimal transmission of objects, the SOS scheduler was given priorities with each object based on whether or not this object was required for DOM rendering, and based on this information, SOS would sort objects by priority as they arrived. Also, it would preempt objects currently being transmitted only if objects of higher priority had arrived on a different connection, as it cannot preempt on the same connection without violating HTTP semantics. The priority information is often given as part of the HTTP header format in HTTP/2 [22]. Additionally, the scheduler considered whether an HTTP object used chunked-encoding and based on that, each packet in the object was treated as a separate object. This is based on the observation that the HTML code of each web page was always sent using chunked-encoding, since each line in the HTML is parsed as soon as it is available and can potentially generate a request for another object, such as CSS files or images. Delay in any particular packet of the HTML can then cause the request to be delayed.

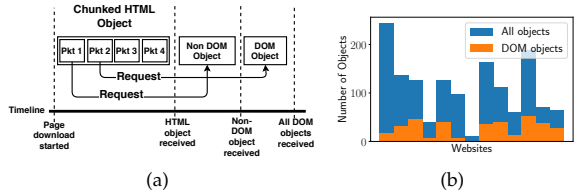

(a)             (b)

Fig. 15: (a) Illustrating the transmission of an example web page with 3 objects, and 2 of which belong to the DOM and are required before rendering can begin. (b) Showing the number of DOM and non-DOM objects in each of the 13 most-visited websites in Ireland

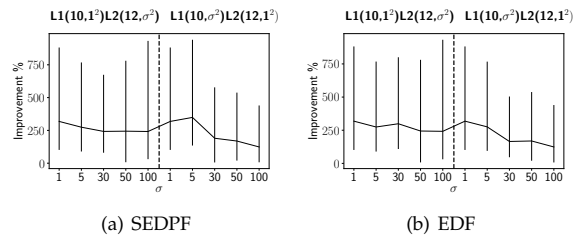

(a) SEDPF        (b) EDF

Fig. 16: Comparing SOS against SEDPF and EDF when downloading popular web pages. Plot shows the improvement in DOM rendering time by SOS, with the line indicating the average and the whiskers indicating the 25th and 95th percentiles.

Most websites tested employed multiple parallel TCP connections for loading the page. Some websites having as many as 10 parallel TCP connections. This is relevant because an HTTP server can use priorities to order the sending of objects across a single connection but it cannot do the same when more than one connection is used as the order of scheduling of the different connections is at the discretion of the transport layer.

Figure 16 plots the measured delay performance of the SOS scheduler compared to that when using the SEDPF and EDF schedulers for a range of path conditions. Since the performance of all schedulers is evaluated for the same measured object traces, so the difference in performance is due to differences in scheduling decisions. The whiskers on the graphs indicate the 25th and 95th percentile performance gains, while the line indicates the mean gain. It can be seen that the SOS scheduler yields consistently lower page load times, with the mean improvement of around 2-3x being substantial.

## 6 CONCLUSIONS

In this paper we consider the task of scheduling packet transmissions amongst multiple paths with uncertain, time-varying delay. We make the observation that the requirement is usually to transmit application layer objects (web pages, images, video frames etc) with low latency, and so it is the object delay rather than the per packet delay which is important. This has fundamental implications for multipath scheduler design. We introduce SOS (Stochastic Object-aware Scheduler), the first multipath scheduler that considers application layer object sizes and their relationship to link uncertainty. We demonstrate that SOS reduces the 95% percentile object delivery delay by 50-100% over production WiFi and LTE links compared to state-of-the art schedulers. We extend SOS to utilize FEC and to handle the scheduling multiple objects in parallel. We show that judicious priority scheduling of HTTP objects can lead to a 2-3x improvement in page load times.

## REFERENCES

[1] "Next Generation Protocols – Market Drivers and Key Scenarios," *European Telecommunications Standards Institute (ETSI)*, 2016. [Online]. Available: http://www.etsi.org/images/files/ETSIWhitePapers/etsi_wp17_Next_Generation_Protocols_v01.pdf

[2] A. Wilk, J. Iyengar, I. Swett, and R. Hamilton, "QUIC: A UDP-Based Secure and Reliable Transport for HTTP/2." [Online]. Available: https://tools.ietf.org/html/draft-tsvwg-quic-protocol-00

[3] M. Kim, J. Cloud, A. ParandehGheibi, L. Urbina, K. Fouli, D. J. Leith, and M. Médard, "Congestion control for coded transport layers," in *2014 IEEE International Conference on Communications (ICC)*, Jun. 2014, pp. 1228–1234, ctcp14.

[4] "Open Fast PAth," 2016. [Online]. Available: https://openfastpath.org/

[5] "Commercial usage of Multipath TCP — MPTCP." [Online]. Available: http://blog.multipath-tcp.org/blog/html/2015/12/25/commercial_usage_of_multipath_tcp.html [last accessed 01-August-2018].

[6] G. Sarwar, R. Boreli, E. Lochin, and A. Mifdaoui, "Performance evaluation of multipath transport protocol in heterogeneous network environments," in *2012 International Symposium on Communications and Information Technologies (ISCIT)*, Oct. 2012, pp. 985–990, mptcpEvaluation.

[7] A. Alheid, D. Kaleshi, and A. Doufexi, "Performance Evaluation of MPTCP in Indoor Heterogeneous Networks," in *Proceedings of the 2014 First International Conference on Systems Informatics, Modelling and Simulation*, ser. SIMS '14. Washington, DC, USA: IEEE Computer Society, 2014, pp. 213–218, mptcpInHeterogenousNetworks. [Online]. Available: http://dx.doi.org/10.1109/SIMS.2014.40

[8] Q. De Coninck, M. Baerts, B. Hesmans, and O. Bonaventure, "A First Analysis of Multipath TCP on Smartphones," Mar. 2016, pp. 57–69, firstLookAtMptcpOnSmartphones.

[9] B. Partov and D. J. Leith, "Experimental Evaluation of Multipath Schedulers for LTE/Wi-Fi Devices," in *Proceedings of the Tenth ACM International Workshop on Wireless Network Testbeds, Experimental Evaluation, and Characterization*, ser. WiNTECH '16. New York, NY, USA: ACM, 2016, pp. 41–48, wifiLteMeasurements. [Online]. Available: http://doi.acm.org/10.1145/2980159.2980169

[10] E. Rescorla <ekr@networkresonance.com>, "The Transport Layer Security (TLS) Protocol Version 1.2." [Online]. Available: https://tools.ietf.org/html/rfc5246

[11] C. Raiciu, C. Paasch, S. Barre, A. Ford, M. Honda, F. Duchene, O. Bonaventure, and M. Handley, "How Hard Can It Be? Designing and Implementing a Deployable Multipath TCP," in *Proceedings of the 9th USENIX Conference on Networked Systems Design and Implementation*, ser. NSDI'12. Berkeley, CA, USA: USENIX Association, 2012, pp. 29–29, mptcp. [Online]. Available: http://dl.acm.org/citation.cfm?id=2228298.2228338

[12] S. Ferlin, O. Alay, O. Mehani, and R. Boreli, "BLEST: Blocking estimation-based MPTCP scheduler for heterogeneous networks," in *2016 IFIP Networking Conference (IFIP Networking) and Workshops*, May 2016, pp. 431–439, bLEST.

[13] G. Sarwar, R. Boreli, E. Lochin, A. Mifdaoui, and G. Smith, "Mitigating Receiver's Buffer Blocking by Delay Aware Packet Scheduling in Multipath Data Transfer," in *2013 27th International Conference on Advanced Information Networking and Applications Workshops*, Mar. 2013, pp. 1119–1124, dAP.

[14] F. Yang, Q. Wang, and P. D. Amer, "Out-of-Order Transmission for In-Order Arrival Scheduling for Multipath TCP," in *2014 28th International Conference on Advanced Information Networking and Applications Workshops*, May 2014, pp. 749–752, oTIAS.

[15] K. Chebrolu and R. Rao, "Communication using multiple wireless interfaces," in *2002 IEEE Wireless Communications and Networking Conference Record. WCNC 2002 (Cat. No.02TH8609)*, vol. 1, Mar. 2002, pp. 327–331 vol.1, eDPF.

[16] Y.-s. Lim, E. M. Nahum, D. Towsley, and R. J. Gibbens, "ECF: An MPTCP Path Scheduler to Manage Heterogeneous Paths," in *Proceedings of the 2017 ACM SIGMETRICS / International Conference on Measurement and Modeling of Computer Systems*, ser. SIGMETRICS '17 Abstracts. New York, NY, USA: ACM, 2017, pp. 33–34, eCF. [Online]. Available: http://doi.acm.org/10.1145/3078505.3078552

[17] F. Yang, P. Amer, and N. Ekiz, "A Scheduler for Multipath TCP," in *2013 22nd International Conference on Computer Communication and Networks (ICCCN)*, Jul. 2013, pp. 1–7, a scheduler for mptcp.

[18] D. Ni, K. Xue, P. Hong, and S. Shen, "Fine-grained Forward Prediction based Dynamic Packet Scheduling Mechanism for multipath TCP in lossy networks," in *2014 23rd International Conference on Computer Communication and Networks (ICCCN)*, Aug. 2014, pp. 1–7, fp2s.

[19] A. Garcia-Saavedra, M. Karzand, and D. J. Leith, "Low Delay Random Linear Coding and Scheduling Over Multiple Interfaces," *IEEE Transactions on Mobile Computing*, vol. 16, no. 11, pp. 3100–3114, Nov. 2017.

[20] C. Bovy, H. Mertodimedjo, H. Uijterwaal, P. Van Mieghem, and G. Hooghiemstra, "Analysis of end-to-end delay measurements in Internet," Jan. 2002, analysisOfEndToEndDelay.

[21] M. Karakas, "DETERMINATION OF NETWORK DELAY DISTRIBUTION OVER THE INTERNET," Master Thesis, THE MIDDLE EAST TECHNICAL UNIVERSITY, 2003. [Online]. Available: https://etd.lib.metu.edu.tr/upload/1223155/index.pdf

[22] M. Belshe, M. Thomson, and R. Peon, "Hypertext Transfer Protocol Version 2 (HTTP/2)." [Online]. Available: https://tools.ietf.org/html/rfc7540

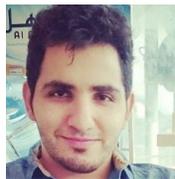

**Kariem Fahmi** obtained his BSc in Computer Science with high distinction from Qatar University in 2015. Kariem worked for a period of two years as a Research Analyst at Bell Labs in Ireland, where he received a certification of outstanding achievement for his contribution to Bell Labs multi-connectivity innovation. He has filed over 10 patents related to multi-path connectivity and network optimization. He is currently a PhD student at Trinity College Dublin under the supervision of Prof. Doug Leith.

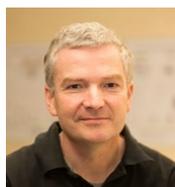

**Doug Leith** graduated from the University of Glasgow in 1986 and was awarded his PhD, also from the University of Glasgow, in 1989. In 2001, Prof. Leith moved to the National University of Ireland, Maynooth and then in Dec 2014 to Trinity College Dublin to take up the Chair of Computer Systems in the School of Computer Science and Statistics. His current research interests include wireless networks, network congestion control, distributed optimization and data privacy.

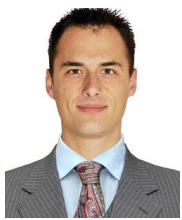

**Stepan Kucera** is a Senior Research Scientist at Bell Laboratories Dublin, Ireland since 2011. He received his Ph.D. degree in Informatics from the Graduate School of Informatics at Kyoto University, Kyoto, Japan, in 2008. Dr. Kucera filed/holds over 50 patents related to 3G and 4G mobile networks, as well as proprietary wireless technologies. He is also the recipient of the 2018 Irish Laboratory Scientist of the Year.

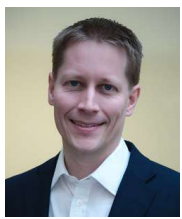

**Holger Claussen** is leader of the Small Cells Research at Bell Labs with a team in Ireland and the US. Dr. Claussen received his Ph.D. degree in signal processing for digital communications from the University of Edinburgh, United Kingdom in 2004. He is author of more than 90 publications 45 granted patent families and further 70 filed patent applications. He is fellow of the World Technology Network, senior member of the IEEE, and member of the IET.